\documentclass[11pt]{article}
 
\usepackage[utf8]{inputenc}
\usepackage[english]{babel}
\usepackage{graphicx}
\usepackage{subcaption}
\usepackage{epsfig}
\usepackage{amssymb}
\usepackage{bm}
\usepackage{amsmath}
\usepackage{color}
\usepackage{hyperref}
\usepackage{times}
\usepackage{float}
\usepackage{times}
\usepackage{physics}
\usepackage{authblk}

\def\be{\begin{equation}}
\def\te{\end{equation}}
\def\bea{\begin{eqnarray}}
\def\nn{\nonumber\\}
\def\tea{\end{eqnarray}}

\begin{document}
\title{A First-Principles Thermodynamic Uncertainty Relation for Shortcuts to Adiabaticity}
\author[1,2]{Guillermo Ezequiel Perna\thanks{gperna@df.uba.ar}}
\author[3]{Federico Centrone\thanks{federico.centrone@icfo.eu}}
\author[1,2]{Esteban Calzetta\thanks{calzetta@df.uba.ar}}
\affil[1]{Departamento de Física, Facultad de Ciencias Exactas y Naturales, Universidad de Buenos Aires,
Ciudad Universitaria, Ciudad de Buenos Aires CP 1428, Argentina}
\affil[2]{Instituto de Física de Buenos Aires (IFIBA), CONICET--Universidad de Buenos Aires,
Ciudad de Buenos Aires CP 1428, Argentina}
\affil[3]{ICFO-Institut de Ciencies Fotoniques, The Barcelona Institute of Science and Technology,
Av. Carl Friedrich Gauss 3, 08860 Castelldefels, Barcelona, Spain}

\maketitle

\begin{abstract}
We study the fundamental limitations of implementing time-dependent Hamiltonian protocols when ``time'' is provided by a quantum clock rather than an external classical parameter. For a parametric harmonic oscillator controlled through a shortcut-to-adiabaticity (STA) schedule and coupled to a minimal clock degree of freedom, tracing out the clock yields an effective reduced dynamics that is a mixture of unitary Gaussian trajectories. Within a noise-dominated regime, we compute the energetic deviation from the target STA outcome and its fluctuations, together with the fidelity to the target evolution and the purity loss of the reduced state, for vacuum and coherent initial states. Combining these observables produces a thermodynamic-uncertainty-type tradeoff that links achievable precision to an irreducible loss of purity set by the clock precision and the protocol sensitivity.
\end{abstract}


\section{Introduction}

Time-dependent Hamiltonians are the standard language of quantum control, quantum thermodynamic strokes, and gate implementation: one prescribes a schedule $H(t)$ and computes the corresponding unitary
$
U=\mathcal{T}\exp\!\left[-\frac{i}{\hbar}\int_{0}^{t_f}\!dt\,H(t)\right].
$
Within this framework, a large body of work has developed protocols that suppress nonadiabatic excitations and enable high-fidelity state preparation in finite time. A prominent example is the family of \emph{shortcuts to adiabaticity} (STA), where suitably engineered drivings reproduce (or closely approximate) the outcome of an adiabatic process without requiring slow evolution \cite{GueryOdelin2019,Torrontegui2013}.

This standard control-theoretic picture treats $t$ as an external classical parameter. However, if one insists on a first-principles description of a closed quantum world, the fundamental dynamics is generated by a time-independent Hamiltonian on a larger Hilbert space. In that setting, a time-dependent Hamiltonian for a subsystem can only emerge effectively: the subsystem must be coupled to additional degrees of freedom whose evolution is used as a time reference. This is not a merely interpretational point. Once a clock is included, the global evolution remains unitary, but the reduced dynamics of the controlled subsystem need not be unitary because the clock can become correlated with it. The operational question is then unavoidable:
\begin{quote}
\emph{What is the intrinsic decoherence and precision limit of a time-dependent unitary protocol when the only source of ``time'' is a quantum clock?}
\end{quote}
 This perspective is central to relational approaches to time in quantum theory \cite{PageWootters1983,GambiniPortoPullin2004}, where dynamics can be understood conditionally with respect to an internal time reference, and it also appears, in a different guise, in proposals where the \emph{quantumness} of clocks leads to an effective loss of coherence for subsystems when the clock is ignored \cite{GambiniPullin2020}.

Because shortcuts to adiabaticity achieve target states in finite time, they are the natural playground for thermodynamic resource analysis in table-top experiments and out-of-equilibrium situations. As long as the evolution is unitary and the control is perfect, these protocols involve no net energetic investment. The actual implementation through a clock-dependent potential is of a quantum nature because so is the variable that correlates to time (the ``hands of the clock''), which is therefore subject to quantum fluctuations \cite{Duncan2025,Mayato2008}. In the case of a system interacting with a heat bath, it is known that statistical fluctuations in the environment lead to thermodynamic uncertainty relations (TURs). We show that the quantum fluctuations in the ``hands of the clock'' lead to limitations in the performance of the adiabatic shortcut in many ways analog to a TUR.

In an autonomous description, these effects do not rely on adding ad hoc classical noise or on continuously measuring the clock. Rather, they arise because a time reference is itself a quantum system whose degrees of freedom become entangled with the controlled dynamics when they generate the schedule and are subject to the uncertainty principle.

The idea of implementing a control protocol through a quantum variable is not new \cite{Mayato2008}. Similar approaches have been explored in various contexts, such as the analysis of errors induced by quantum fluctuations in quantum circuits \cite{Loughlin2023} and to engineer a time-dependent potential by letting an ion fall through a tapered trap (see for example \cite{Myers2022,Abah2012}). The underlying principle in these implementations is that an effective clock is an observable whose average observed value evolves linearly in time with little to no dispersion.

From the reduced point of view in which the clock is discarded, this unavoidable entanglement manifests as a spread of effective ``histories'' for the control and hence as an effectively nonunitary map on the system. In other words, even in the absence of any environment, can the \emph{implementation} of a time-programmed protocol generate mixedness and limit fidelity to the intended target, solely because timekeeping is quantum?

The present work pursues this question in a setting where both the control objective and the clock-induced limitation can be characterized analytically. We consider a parametric harmonic oscillator driven by a STA protocol and replace the externally prescribed time dependence by a fully autonomous system--clock model in which the oscillator frequency depends on a clock coordinate. This builds directly on Ref.~\cite{NotQuiteFreeSTA2018}, where an autonomous clock implementation of STA was introduced and it was shown that, even in the absence of any external environment, system--clock correlations generically deflect the evolution from the ideal STA trajectory and lead to a spread of outcomes once the clock is ignored. 

Here we go beyond that analysis by reframing the effect as a limitation of \emph{time-programmed quantum operations}: we treat the ideal STA evolution (obtained in the limit of a sharply defined clock trajectory) as a \emph{target unitary/state-preparation task}, and we quantify (i) the degradation of the target (fidelity), (ii) the induced mixedness upon tracing out the clock (purity loss/irreversibility), and (iii) energetic deviations and fluctuations as complementary diagnostics of the same mechanism. Importantly, the resulting “noise” is not an extra stochastic ingredient added by hand: it is a compact parametrization of the unitary entangling dynamics between system and a quantum time reference, viewed from the reduced description in which the clock degrees of freedom are discarded.

STA provides a particularly clean benchmark for this purpose. In the ideal description with a perfectly classical time parameter, STA can be arranged so that deterministic nonadiabatic excitations vanish by construction. Any residual excitations, loss of fidelity, or loss of purity in our autonomous model can therefore be attributed to the quantumness of the time reference---rather than to an imperfect choice of schedule---thereby isolating what is truly \emph{intrinsic to timekeeping}.

Methodologically, we start from the unitary dynamics of the combined system--clock and trace out the clock using a Feynman--Vernon influence functional \cite{FeynmanVernon1963,FeynmanHibbs}. In the regime studied here, the clock induces an effectively stochastic modulation of the oscillator frequency: each realization corresponds to a unitary evolution of the oscillator, while decoherence arises only after averaging over realizations (equivalently, after discarding the clock). This provides a controlled and microscopic route from an autonomous Hamiltonian description to an operationally meaningful “error model” for time-dependent control.

Conceptually, we aim to connect this clock-induced mechanism to a tradeoff between \emph{precision} and \emph{irreversibility} for time-programmed protocols. Such tradeoffs resonate with broader discussions of the thermodynamic cost of timekeeping and the limitations of clocks \cite{Pearson2021TimekeepingCost,VanVuSaito2025}, as well as recent developments on thermodynamic uncertainty relations and precision--cost tradeoffs in driven systems \cite{Farina2024}. In our case, the relevant figures of merit are not extracted from an external bath or measurement record: they arise from unitary system--clock correlations alone.

The quantum nature of the ``control knob'' has long been recognized as a potential limitation for idealized time-dependent protocols: quantizing the control field (e.g.,\ a laser mode) generically entangles it with the target and induces dephasing/errors when the control is discarded \cite{vanEnkKimble2002,GeaBanacloche2002,ChanSham2011,SilberfarbDeutsch2004}, with error scalings tied to control resources \cite{GeaBanaclocheMiller2008,Milburn2012}. Complementary approaches model imperfect timekeeping as an uncertainty in the duration of an intended operation (random time) \cite{XuerebEtAl2023,Neri2025}, while recent work develops fully autonomous clock-driven dynamics in which a quantum clock generates the schedule without external time dependence \cite{MalabarbaShortKammerlander2015,LautenbacherEtAl2025}. Our contribution lies at this intersection: we study a genuinely time-programmed Hamiltonian protocol (STA for a parametric oscillator) implemented by a minimal quantum clock, derive the induced random-unitary description from first principles, and—building on the autonomous-STA framework of \cite{NotQuiteFreeSTA2018}—connect fidelity loss and purity loss to energetic deviations through a thermodynamic-uncertainty-type tradeoff, with analytic results in a controlled small-noise regime (including arbitrary Gaussian inputs).

The paper is organized as follows. In Sec.~\ref{sec:model}, we introduce the autonomous system--clock Hamiltonian and derive the reduced dynamics of the oscillator by tracing out the clock degrees of freedom. In Sec.~\ref{sec:dynamics}, we analyze the unitary dynamics associated with a single clock trajectory and express the evolution in terms of symplectic maps and Bogoliubov coefficients. In Sec.~\ref{sec:observables}, we define the relevant observables —energetic deviations, fidelity with respect to the target STA evolution, and purity loss— and evaluate their clock-averaged behavior in the small-noise regime. These quantities are then combined to establish a thermodynamic-uncertainty-type relation linking precision, energetic fluctuations, and irreversibility. In Sec.~\ref{sec:numerics}, we present numerical results for representative driving protocols. Finally, Sec.~\ref{sec:conclusions} summarizes our conclusions. Technical derivations and additional details are collected in the Appendices.

% we express energetic deviations and fluctuations relative to the ideal STA target in terms of Bogoliubov coefficients. 
% In Sec.~\dots\ we compute the fidelity to the target and the purity loss of the reduced oscillator state, and we combine these quantities into a thermodynamic-uncertainty-type tradeoff linking precision and irreversibility.

% ============================================================
%  MODEL SECTION (MAIN TEXT)
%  Replace your current \section*{Preliminaries} block by this
% ============================================================

\section{Autonomous clock model and reduced dynamics}
\label{sec:model}

\subsection{Target protocol and STA benchmark}
\label{subsec:sta-benchmark}

We consider a parametric harmonic oscillator (system) with canonical variables $(x,p)$ and mass $m$. In the standard control description, one prescribes a time-dependent frequency schedule $\omega(t)$ and studies the unitary generated by
\be
H_{\rm bare}(t)=\frac{p^2}{2m}+\frac{m}{2}\,\omega^2(t)\,x^2 .
\label{eq:Hbare}
\te
The time-dependent quantum harmonic oscillator provides a canonical setting to study the transition between adiabatic and sudden driving regimes \cite{Martinez-Tibaduiza_2021}. For a given schedule $\omega(t)$, a shortcut-to-adiabaticity (STA) protocol consists in engineering a modified driving frequency $\Omega(t)$ such that, with $t$ treated as an external classical parameter, the evolution reproduces the desired adiabatic mapping in finite time. In the present manuscript, we use the familiar STA construction
\be
H_{\rm STA}(t)=\frac{p^2}{2m}+\frac{m}{2}\,\bar\Omega^2(t)\,x^2,
\label{eq:HSTA}
\te
with
\be
\bar\Omega^2(t)=\omega^2(t)+\frac{1}{2}\frac{\ddot\omega(t)}{\omega(t)}-\frac{3}{4}\left(\frac{\dot\omega(t)}{\omega(t)}\right)^2 .
\label{eq:Omegabar}
\te
Similar STA constructions arise in other systems whose dynamics reduce to harmonic oscillators with time-dependent frequencies, such as electromagnetic modes in cavities with moving boundaries \cite{e25010018}.

We assume that the driving is well-behaved at the endpoints (in particular, first and second derivatives vanish at $t=t_i$ and $t=t_f$) so that the initial and final instantaneous frequencies are unambiguously defined. In what follows, the STA evolution associated with $\bar\Omega(t)$ will serve as the \emph{target} protocol.

\subsection{Autonomous implementation with a quantum clock}
\label{subsec:autonomous-clock}

To formulate time dependence from first principles, we replace the external parameter $t$ by an explicit clock degree of freedom. The clock is modeled as a single pointer coordinate $X$ with conjugate momentum $P$, evolving freely with mass $M$. The autonomous system--clock Hamiltonian is taken to be
\be
H=\frac{p^2}{2m}+\frac{m}{2}\,\Omega^2[X]\,x^2+\frac{P^2}{2M}.
\label{eq:Htot}
\te
Here $\Omega^2[X]$ is a smooth function of the clock coordinate, chosen so that along a prescribed mean clock trajectory $\bar X(t)$ the system experiences the target STA schedule,
\be
\Omega^2[\bar X(t)]=\bar\Omega^2(t).
\label{eq:targettrajectory}
\te
We will assume an initially uncorrelated state at $t=t_i$,
\be
\rho_{\rm tot}(t_i)=\rho_S(t_i)\otimes\rho_C(t_i),
\label{eq:initialproduct}
\te
and define the reduced state of the oscillator at $t_f$ by tracing out the clock,
\be
\rho_S(t_f)=\Tr_C\!\left[U(t_f,t_i)\,\rho_{\rm tot}(t_i)\,U^\dagger(t_f,t_i)\right].
\label{eq:reducedstate}
\te
The global evolution is unitary; any effective nonunitarity for the oscillator arises solely because the clock is discarded.

\subsection{Influence functional representation}
\label{subsec:influencefunctional-main}

The reduced dynamics may be expressed with a Feynman--Vernon influence functional \cite{FeynmanVernon1963,FeynmanHibbs}. In coordinate representation, one can write
\bea
\rho_S(x_f,x_f',t_f)
&=&\int dx_i\,dx_i' \int\!Dx\,Dx'\;
e^{\frac{i}{\hbar}\left(S_S[x]-S_S[x']\right)}\,e^{\frac{i}{\hbar}S_{IF}[x,x']}\,
\rho_S(x_i,x_i',t_i),
\label{eq:rhoS_FV_compact}
\tea
where $S_S$ is the system action evaluated along the two paths, and $S_{IF}$ encodes the effect of the clock after it is traced out. The full derivation (including the explicit path-integral representation for the trace over clock histories and the resulting expression for $S_{IF}$) is given in Appendix~\ref{app:influence-functional}.

We now expand the clock coordinate around a prescribed mean trajectory,
\be
X(t)=\bar X(t)+\xi(t),
\label{eq:Xsplit}
\te
and retain the leading (linear) response of the frequency to clock fluctuations,
\be
\Omega^2[X(t)] \simeq \Omega^2[\bar X(t)] + \left.\frac{d\Omega^2}{dX}\right|_{\bar X(t)} \xi(t)
\equiv \bar\Omega^2(t)+V(t)\,\xi(t),
\label{eq:linearresponse}
\te
where
\be
V(t)=\left.\frac{d\Omega^2}{dX}\right|_{\bar X(t)} .
\label{eq:Vdef}
\te
At this order, the influence functional takes the standard Gaussian form \cite{CalzettaHu}
\be
S_{IF}[x,x']=S_D[x,x']+i\,S_N[x,x'],
\label{eq:SIFsplit}
\te
with dissipation and noise contributions governed by the clock commutator and anticommutator kernels (definitions and explicit expressions are collected in Appendix~\ref{app:influence-functional} and Appendix~\ref{app:clock-kernels}).

\subsection{Noise-dominated single-parameter reduction}
\label{subsec:singleparameter}

For a free pointer clock, the Heisenberg evolution of $\xi$ is linear,
\be
\xi(t)=\hat\xi+\frac{\hat P}{M}\,t,
\label{eq:xiHeis}
\te
and the clock kernels become simple functions of the clock state (Appendix~\ref{app:clock-kernels}). In particular,
\bea
D(t,t')&=& i\langle[\xi(t),\xi(t')]\rangle\,\theta(t-t')
=\frac{\hbar}{M}(t-t')\,\theta(t-t'),
\label{eq:Dkernel_main}
\\
N(t,t')&=&\langle\{\xi(t),\xi(t')\}\rangle
=2\langle\hat\xi^2\rangle+\frac{2}{M^2}\langle\hat P^2\rangle\,t\,t' .
\label{eq:Nkernel_main}
\tea

In the regime of interest here, we make two simplifications that lead to a compact reduced description:
(i) we focus on clock states for which the term proportional to $\langle\hat P^2\rangle$ dominates the noise kernel, so that the $t t'$ structure of $N(t,t')$ is the leading contribution; and
(ii) we work in a noise-dominated approximation in which the dissipative contribution $S_D$ is neglected at leading order (see Appendix~\ref{app:hs-reduction} for details and the resulting representation).

With these assumptions, the noise functional factorizes as
\be
S_N[x,x']
\simeq
\frac{m^2}{8\hbar}\,\sigma_v^2
\left[\int_{t_i}^{t_f} dt\; t\,V(t)\,\big(x^2(t)-x'^2(t)\big)\right]^2,
\qquad
\sigma_v^2=\frac{\langle\hat P^2\rangle}{M^2}.
\label{eq:SNfactorized_main}
\te
Using a Hubbard--Stratonovich representation (Appendix~\ref{app:hs-reduction}), one obtains
\bea
e^{\frac{i}{\hbar}S_{IF}[x,x']}
&\simeq&
\int dv\,P(v)\,
\exp\!\left[
-\frac{i m v}{2\hbar}\int_{t_i}^{t_f} dt\; t\,V(t)\,\big(x^2(t)-x'^2(t)\big)
\right],
\label{eq:SIF_HS_main}
\tea
where $P(v)$ is a Gaussian measure with $\langle v\rangle=0$ and $\langle v^2\rangle=\sigma_v^2$.

Equation~\eqref{eq:SIF_HS_main} shows that the reduced dynamics can be written as an average over unitary evolutions labeled by a single real parameter $v$. Equivalently,
\be
\rho_S(t_f)\simeq \int dv\,P(v)\;U_v(t_f,t_i)\,\rho_S(t_i)\,U_v^\dagger(t_f,t_i),
\label{eq:randomunitarymap}
\te
where $U_v$ is the oscillator unitary generated by a parametric Hamiltonian with an effective, $v$-dependent frequency
\bea
\Omega^2(t)
&=&\Omega^2[\bar X(t)] + V(t)\,v\,t
\equiv \bar\Omega^2(t)+v\,\delta\Omega^2(t),
\label{eq:Omega_stochastic_main}
\\
\delta\Omega^2(t)&=& t\,V(t).
\label{eq:deltaOmega_main}
\tea
For each fixed $v$ the oscillator evolution is unitary; the reduced state becomes mixed only after averaging over $v$, i.e., after discarding the clock.

\subsection{Assumptions and regime of validity}
\label{subsec:validity}

For clarity, we summarize the assumptions behind Eqs.~\eqref{eq:randomunitarymap}--\eqref{eq:Omega_stochastic_main}:

\begin{itemize}
\item \textbf{Closed global dynamics.} The combined system (oscillator + clock) evolves unitarily; reduced nonunitarity arises only from tracing out the clock, Eq.~\eqref{eq:reducedstate}.
\item \textbf{Minimal clock model.} The clock is represented by a single free pointer degree of freedom with Hamiltonian $P^2/2M$.
\item \textbf{Weak clock fluctuations.} We expand around a prescribed mean trajectory $X(t)=\bar X(t)+\xi(t)$ and retain only the leading (linear) response of $\Omega^2[X]$ to $\xi$, Eq.~\eqref{eq:linearresponse}.
\item \textbf{Noise-dominated approximation.} We neglect the dissipative influence contribution $S_D$ at leading order, keeping only the noise functional $S_N$.
\item \textbf{Single-parameter reduction.} We focus on the dominant $t t'$ contribution in the noise kernel, yielding the factorized form Eq.~\eqref{eq:SNfactorized_main} and the single Gaussian parameter $v$ with variance $\sigma_v^2=\langle\hat P^2\rangle/M^2$.
\item \textbf{Well-defined endpoints.} Protocols are assumed to have well-defined asymptotic frequencies at $t_i,t_f$, so that the target STA evolution is unambiguously defined.
\end{itemize}

\section{Unitary dynamics for single trajectory}
\label{sec:dynamics}

In the noise-dominated regime, the clock-traced output at $t_f$ is a random-unitary mixture
\be
\bar\rho_S(t_f)\simeq \int dv\,P(v)\,\rho_v(t_f),
\qquad 
\rho_v(t_f)=U_v(t_f,t_i)\,\rho_S(t_i)\,U_v^\dagger(t_f,t_i),
\label{eq:rho_average_def}
\te
where for each realization $v$ the oscillator evolves under the quadratic Hamiltonian
\be
H_v(t)=\frac{p^2}{2m}+\frac{m}{2}\Omega_v^2(t)\,x^2,
\qquad 
\Omega_v^2(t)=\bar\Omega^2(t)+v\,\delta\Omega^2(t),
\label{eq:Hv_dyn}
\te
with $\delta\Omega^2(t)=tV(t)$ (Sec.~\ref{subsec:singleparameter}). For each fixed $v$ the evolution is unitary and
Gaussian-preserving; nonunitarity arises only after averaging over $v$.

\subsection{Symplectic evolution of the quadratures}
\label{subsec:symplectic_evolution_main}

Define the quadrature vector and symplectic form
\be
\bm R \equiv (x,p)^{\mathsf T},\qquad 
J=\begin{pmatrix}0&1\\-1&0\end{pmatrix},\qquad [x,p]=i\hbar .
\label{eq:R_J_def}
\te
The Heisenberg equations generated by Eq.~\eqref{eq:Hv_dyn} are linear,
\be
\frac{d}{dt}\bm R(t)=A_v(t)\,\bm R(t),\qquad 
A_v(t)=\begin{pmatrix}
0 & 1/m\\
-\,m\Omega_v^2(t) & 0
\end{pmatrix},
\label{eq:R_Heis_linear}
\te
so there exists a fundamental matrix $S_v(t)$ such that
\bea
\bm R(t)&=&S_v(t)\,\bm R(t_i),\label{eq:R_Sv_def}\\
\frac{d}{dt}S_v(t)&=&A_v(t)\,S_v(t),\qquad S_v(t_i)=\mathbb I_2.\label{eq:Sv_ODE_main}
\tea
For all $t$, $S_v(t)$ is symplectic,
\be
S_v^{\mathsf T}(t)\,J\,S_v(t)=J,
\label{eq:symplectic_condition}
\te
hence $S_v(t)\in Sp(2,\mathbb R)$ and $\det S_v(t)=1$. We write $S_v\equiv S_v(t_f)$ for the final-time map.

Equivalently, $S_v(t)$ can be parametrized by two classical solutions $u_v,w_v$ of
\be
\ddot y+\Omega_v^2(t)\,y=0
\label{eq:classical_eq}
\te
with initial data $u_v(t_i)=1,\dot u_v(t_i)=0$ and $w_v(t_i)=0,\dot w_v(t_i)=1$, giving
\be
S_v(t)=
\begin{pmatrix}
u_v(t) & \frac{1}{m}w_v(t)\\[1mm]
m\dot u_v(t) & \dot w_v(t)
\end{pmatrix}.
\label{eq:Sv_uw_main}
\te
This representation makes transparent how each unitary trajectory acts linearly on $(x,p)$.

\subsection{Symplectic map and Bogoliubov coefficients}
\label{subsec:sv_bogo_main}

For each realization $v$, the oscillator Hamiltonian $H_v(t)$ is quadratic; hence, the Heisenberg evolution of the quadratures
$\bm R=(x,p)^{\mathsf T}$ is linear:
\be
\bm R(t_f)=S_v\,\bm R(t_i),\qquad S_v\in Sp(2,\mathbb R),\qquad S_v^{\mathsf T}J S_v=J,
\label{eq:Sv_main}
\te
with $J=\bigl(\begin{smallmatrix}0&1\\-1&0\end{smallmatrix}\bigr)$.
At the endpoint $t_f$, we introduce the ladder operator associated with the instantaneous frequency
$\omega_f=\bar\Omega(t_f)$,
\be
a_f=\sqrt{\frac{m\omega_f}{2\hbar}}\,x
+i\sqrt{\frac{1}{2\hbar m\omega_f}}\,p,
\label{eq:af_def}
\te
so that $H_f=\hbar\omega_f(a_f^\dagger a_f+\tfrac12)$.
Any symplectic map $S_v$ therefore induces an equivalent Bogoliubov transformation on $a_f$
\cite{CalzettaHu,DiCastroRaimondi2015}, which is an isomorphism of the canonical commutation relation algebra,
\be
U_v^\dagger a_f U_v=\alpha_v\,a_f+\beta_v\,a_f^\dagger,
\qquad |\alpha_v|^2-|\beta_v|^2=1,
\label{eq:bogo_def}
\te
and conversely $(\alpha_v,\beta_v)$ uniquely determine $S_v$.

Writing
$S_v=\bigl(\begin{smallmatrix}a&b\\c&d\end{smallmatrix}\bigr)$ in the $(x,p)$ basis, a direct substitution of
Eq.~\eqref{eq:af_def} into $\bm R\mapsto S_v\bm R$ gives the explicit dictionary
\bea
\alpha_v &=& \frac12\left[a+d+i\left(\frac{c}{m\omega_f}-m\omega_f\,b\right)\right],\label{eq:alpha_from_S}\\
\beta_v  &=& \frac12\left[a-d+i\left(\frac{c}{m\omega_f}+m\omega_f\,b\right)\right],\label{eq:beta_from_S}
\tea
which automatically satisfies $|\alpha_v|^2-|\beta_v|^2=1$ whenever $S_v^{\mathsf T}JS_v=J$.
This identification provides a transparent physical interpretation: $\beta_v$ is the \emph{squeezing/parametric excitation}
generated by the $v$-dependent modulation. In particular, for a vacuum input, one has
$\langle n\rangle_v=\langle a_f^\dagger a_f\rangle_v=|\beta_v|^2$, and the excess final energy is
$\Delta E(v)=\hbar\omega_f|\beta_v|^2$.

In the small-noise regime relevant for the clock-induced mixture, the dependence on $v$ is perturbative,
\be
\beta_v = i v\,\beta_1+O(v^2),
\label{eq:beta_smallv}
\te
so that $\overline{\langle n\rangle}=\int dv\,P(v)\,|\beta_v|^2\simeq \sigma_v^2|\beta_1|^2$.
Hence $|\beta_1|$ quantifies the \emph{susceptibility} of the target protocol to the clock-induced stochastic modulation,
and it is the natural dimensionless control parameter behind the TUR plots shown below.
The construction of $S_v$ and its controlled expansion in $v$ (including how to preserve symplecticity order by order)
is provided in Appendix~\ref{app:Sv_perturb}, together with an explicit consistency check reproducing the vacuum/coherent
Bogoliubov formulas from the symplectic overlap expressions in Appendix \ref{app:equivalence_dictionary}.

\section{Observables, clock averages, and TUR}
\label{sec:observables}

In the noise-dominated regime the reduced output at $t_f$ is the random-unitary mixture
\be
\bar\rho_S(t_f)\simeq \int dv\,P(v)\,\rho_v(t_f),\qquad 
\rho_v(t_f)=U_v\,\rho_S(t_i)\,U_v^\dagger,
\label{eq:rho_bar_def}
\te
with $P(v)$ Gaussian (Sec.~\ref{subsec:singleparameter}). For each fixed $v$, $U_v$ is generated by the quadratic Hamiltonian
Eq.~\eqref{eq:Omega_stochastic_main} and therefore maps Gaussian states to Gaussian states (Sec.~\ref{sec:dynamics}).

In this section, we define the figures of merit used to quantify the deviation from the target STA trajectory ($v=0$),
and we show how their clock averages can be evaluated analytically in a controlled small-noise regime.

% ------------------------------------------------------------
\subsection{Energy deviation and its clock-induced fluctuations}
% ------------------------------------------------------------

As an energetic diagnostic at the endpoint, we use the instantaneous oscillator Hamiltonian
\be
H_f=\frac{p^2}{2m}+\frac{m\omega_f^2}{2}x^2
=\frac12\,\bm R^{\mathsf T}G_f\,\bm R,
\qquad
G_f=\mathrm{diag}\!\left(m\omega_f^2,\frac1m\right),
\label{eq:Hf_def}
\te
where $\omega_f\equiv \bar\Omega(t_f)$ and $\bm R=(x,p)^{\mathsf T}$.

For each realization $v$, we define the \emph{energetic deviation from the target} as
\be
\Delta E(v)\equiv \langle H_f\rangle_v-\langle H_f\rangle_0,
\label{eq:DeltaE_def}
\te
and its clock average
\be
\overline{\Delta E}\equiv \int dv\,P(v)\,\Delta E(v).
\label{eq:DeltaE_bar_def}
\te
We quantify energetic ``fluctuations'' in two distinct senses. First, each unitary trajectory $\rho_v$ generally
has nonzero \emph{quantum} energy uncertainty at the endpoint. We therefore define the excess (clock-induced)
quantum variance
\be
\sigma_E^2(v)\equiv \text{Var}_{\rho_v}(H_f)-\text{Var}_{\rho_0}(H_f),
\qquad 
\overline{\sigma_E^2}\equiv \int dv\,P(v)\,\sigma_E^2(v),
\label{eq:sigmaE_def}
\te
which vanishes in the ideal target evolution and isolates the additional uncertainty produced by the clock.

Second, the clock also induces a \emph{classical} spread across trajectories in the mean energy,
\be
\text{Var}\!\big(\langle H_f\rangle_v\big)\equiv 
\int dv\,P(v)\,\langle H_f\rangle_v^{\,2}
-\left(\int dv\,P(v)\,\langle H_f\rangle_v\right)^2,
\label{eq:Varv_meanE}
\te
which we use as an auxiliary diagnostic (Appendix~\ref{app:gaussian_averages}).
For Gaussian inputs, both $\langle H_f\rangle_v$ and $\text{Var}_{\rho_v}(H_f)$ admit closed expressions in terms of
the output first and second moments (Appendix~\ref{app:gaussian_averages}).

% ------------------------------------------------------------
\subsection{Irreversibility and target overlap }
% ------------------------------------------------------------

In our setting, the global evolution (system + clock) is unitary, and the only source of irreversibility for the
controlled system is the coarse-graining associated with discarding the clock. In the noise-dominated regime, the
reduced output at $t_f$ is the random-unitary mixture
\be
\bar\rho \equiv \int dv\,P(v)\,\rho_v,\qquad \rho_v \equiv U_v\,\rho_i\,U_v^\dagger,
\label{eq:rho_bar_def_main}
\te
and we take $\rho_0\equiv\rho_{v=0}$ as the target output.

A natural information-theoretic measure of mixing is the purity
\be
\mathcal P \equiv \Tr(\bar\rho^{\,2}),\qquad
\mathcal P_0 \equiv \Tr(\rho_0^{\,2}),
\label{eq:purities_def}
\te
and we define the associated R\'enyi-2 entropy \cite{Renyi1961} $S_2(\rho)\equiv -\ln\Tr(\rho^2)$ and its increment
\be
\Delta S_2 \equiv S_2(\bar\rho)-S_2(\rho_0)
= -\ln\!\left(\frac{\mathcal P}{\mathcal P_0}\right)\ge 0.
\label{eq:DeltaS2_def}
\te
While the von Neumann entropy increase $S(\bar\rho)-S(\rho_0)$ is a natural notion of entropy production, $S_2$ is
particularly convenient here because it is analytically tractable and directly tied to Gaussian overlap formulas.
When the target output $\rho_0$ is pure, $\Delta S_2$ also provides a conservative proxy for the true entropy increase
since $S(\bar\rho)\ge S_2(\bar\rho)$ and $S(\rho_0)=S_2(\rho_0)=0$.

To quantify closeness to the target, we use the normalized Hilbert--Schmidt overlap
\be
F_{\mathrm{HS}}\equiv
\frac{\Tr(\rho_0\,\bar\rho)}{\Tr(\rho_0^2)}
=\frac{1}{\mathcal P_0}\int dv\,P(v)\,\Tr(\rho_0\rho_v),
\label{eq:FHS_def_main}
\te
which coincides with the usual fidelity whenever $\rho_0$ is pure and is especially convenient because it reduces
clock averages to Gaussian integrals over overlaps.

A key point is that purity loss and target overlap are not independent. By Cauchy--Schwarz for the Hilbert--Schmidt
inner product,
\be
\big[\Tr(\rho_0\bar\rho)\big]^2 \le \Tr(\rho_0^2)\,\Tr(\bar\rho^2)=\mathcal P_0\,\mathcal P.
\te
Dividing by $\mathcal P_0^2$ gives the universal bound
\be
F_{\mathrm{HS}}^2 \le \frac{\mathcal P}{\mathcal P_0}
\qquad\Longleftrightarrow\qquad
-\,2\ln F_{\mathrm{HS}} \ \ge\ \Delta S_2,
\label{eq:TUR_general}
\te
which expresses a model-independent tradeoff: achieving large overlap with the target output requires the
clock-traced state to remain nearly pure.

% ------------------------------------------------------------
\subsection{Closed clock averages in the small-noise regime}
% ------------------------------------------------------------

When the relevant weight of $P(v)$ lies in the regime where a quadratic expansion of the log-overlap kernel is accurate,
the $v$-integrals for $\mathcal P$ and $F_{\mathrm{HS}}$ can be performed analytically.
Writing $\sigma_v^2\equiv\langle v^2\rangle$ and denoting by $a,b,c$ the protocol- and input-dependent coefficients
defined in Appendix~\ref{app:gaussian_averages}, one finds
\bea
\mathcal P 
&\simeq&
\frac{\mathcal P_0}{
\sqrt{(1+a\sigma_v^2)^2-(c\sigma_v^2)^2}
}\;
\exp\!\left[
\frac{b^2\sigma_v^2\big(1+(a+c)\sigma_v^2\big)}{(1+a\sigma_v^2)^2-(c\sigma_v^2)^2}
\right],
\label{eq:P_closed_main}
\\
F_{\mathrm{HS}}
&\simeq&
\frac{1}{\sqrt{1+a\sigma_v^2}}\;
\exp\!\left(\frac{b^2\sigma_v^2}{2[1+a\sigma_v^2]}\right).
\label{eq:FHS_closed_main}
\tea
In many situations (in particular when the target $v=0$ trajectory is a stationary point of the overlap kernel), one has $b=0$,
and \eqref{eq:P_closed_main}--\eqref{eq:FHS_closed_main} reduce to pure square-root forms.

Finally, the energetic deviation also scales as $\overline{\Delta E}=\chi_E\,\sigma_v^2+O(\sigma_v^4)$, with an explicit
susceptibility $\chi_E$ given in Appendix~\ref{app:gaussian_averages}.
Eliminating $\sigma_v^2$ between $\overline{\Delta E}$ and $\Delta S_2$ yields
\be
\Delta S_2 = \frac{a}{\chi_E}\,\overline{\Delta E}\;+\;O(\sigma_v^4),
\label{eq:DeltaS2_energy_link}
\te
so that the universal inequality Eq.~\eqref{eq:TUR_general} becomes, in the controlled small-noise regime,
\be
-\,\ln F_{\mathrm{HS}}\ \ge\ \frac12\,\Delta S_2
\ \simeq\ 
\frac{a}{2\chi_E}\,\overline{\Delta E},
\label{eq:TUR_final_main}
\te
which is a geometric precision-irreversibility tradeoff:
improving precision (larger overlap with the target) requires suppressing the irreversibility induced by discarding the clock,
and this same irreversibility controls the energetic deviation from the target STA outcome.

\paragraph{TUR-type tradeoff.}
Beyond the universal precision--irreversibility inequality Eq.~\eqref{eq:TUR_general}, we also introduce a
thermodynamic-uncertainty-type figure of merit that combines (i) irreversibility due to discarding the clock and
(ii) energetic precision at the endpoint. We use the linear-entropy production
\be
\mathcal S_L \equiv 1-\frac{\mathcal P}{\mathcal P_0}\ge 0,
\label{eq:SL_def}
\te
and define the TUR ratio (see Appendix \ref{app:TUR})
\be
\mathcal R_E \equiv 
\mathcal S_L\,\frac{\overline{\sigma_E^2}}{\overline{\Delta E}^{\,2}}.
\label{eq:RE_def}
\te
In the noise-dominated small-noise regime (where the random-unitary reduction is controlled), we find that
$\mathcal R_E$ approaches a protocol-independent constant for the benchmark families studied below
(vacuum and displaced vacuum), namely
\be
\mathcal R_E \ \ge\ 2\;+\;O(\sigma_v^2),
\label{eq:TUR_energy_main}
\te
for the vacuum state and
\be
\mathcal R_E \ \ge\ f\left(\mu\right)\;+\;O(\sigma_v^2),
\label{eq:TUR_energy_main_mu}
\te
for coherent states with annihilation operator eigenvalue $\mu$, where $f\left(\mu\right)$ is an asymptotically decreasing function in $\left|\mu\right|$ (see Appendix \ref{app:TUR}), with saturation at leading order as $\sigma_v^2\to 0$.
Equations~\eqref{eq:TUR_energy_main} and~\eqref{eq:TUR_energy_main_mu} are the TUR statements we test numerically and plot in the following section.

The studied states, being initially pure coherent states, are non-thermal (except for the vacuum) and have no corresponding temperature associated. After the evolution, they lose purity but remain non-thermal and out-of-equilibrium. A closely related quantity to an ``effective temperature'' is $T_{\mathrm{eff}}\equiv \hbar\omega\langle n\rangle/k_B=\hbar\omega\left|\mu\right|^2/k_B$, which is the mean value of energy. With this analogy, our result may be seen to reproduce previous ones, for example Eq.~(5) of Ref.~\cite{HorowitzGingrich2020} in the large $T_{\mathrm{eff}}$ limit.

\section{Numerical tests and examples}
\label{sec:numerics}
We studied our model for two different frequency evolution protocols, one standard in toy-models for table-top experiments \cite{FinProt1,FinProt2} and the other borrowed from cosmological applications \cite{InfProt1,InfProt2}. We call them ``finite'' and ``infinite'' time protocols respectively.
\subsection{Finite time protocol}
\begin{figure}[H]
    \centering
    \includegraphics[width=0.7\linewidth]{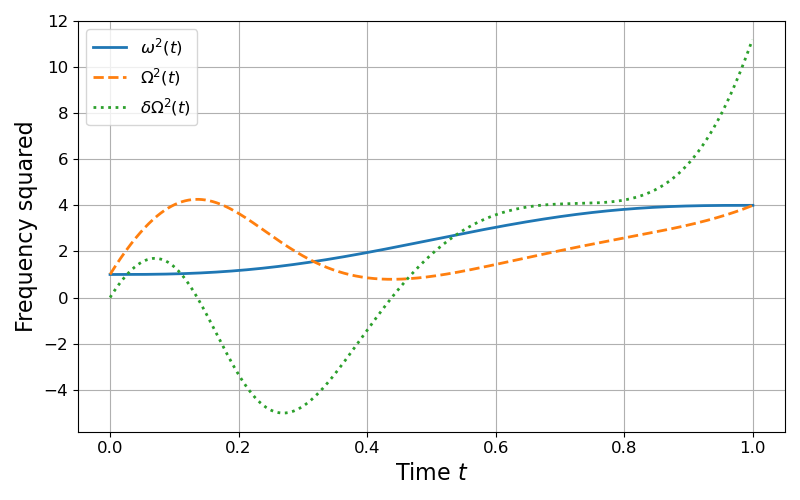}
    \caption{Frequency protocols for the finite-time driving schedule. The blue solid curve shows the prescribed frequency squared $\omega^2\left(t\right)$ defined in Eq.~\eqref{eq:fin_prot}, which smoothly interpolates between the initial value $\omega_i$ and the final value $\omega_f$ during a total duration $\tau$. The orange dashed curve shows the corresponding shortcut-to-adiabaticity (STA) frequency $\bar\Omega^2 \left( t \right)$ constructed from Eq.~\eqref{eq:Omegabar}, which guarantees that the oscillator follows the adiabatic mapping in finite time when time is treated as an external parameter. The green dotted curve shows the STA deviation $\delta \Omega^2$ as defined in Eq.~\eqref{eq:deltaOmega_main}. Frequencies are shown in units where $\omega_i = 1$ and $\omega_f = 2$, and time is measured in units of the protocol duration $\tau$.}
    \label{Fig:finite_time}
\end{figure}

The frequency schedule for this case is
\be
\label{eq:fin_prot}
\omega^2(t) = \omega_i^2 + \left( \omega_f^2 - \omega_i^2 \right) \left[ 10 s^3 - 15 s^4 + 6 s^5 \right],
\te
with $s = \frac{t}{\tau}$. This protocol begins at $s = 0$ and ends at $s = 1$, satisfying $\dot{\omega}(0) = \dot{\omega}(1) = 0$ and $\ddot{\omega}(0) = \ddot{\omega}(1) = 0$ and also the Hamiltonian initial and final conditions. The protocol and its STA are plotted in Figure~\ref{Fig:finite_time}.

%Blue line: frequency squared for the finite time protocol. Yellow line: frequency squared for the shortcut to adiabaticity of the finite time protocol. Green line: $\delta \Omega^2$ as defined in Eq. (\ref{eq:deltaOmega_main}). Everything is plotted in units of $\tau$. $\omega_i^2 = 1$, $\omega_i^2 = 2$.

\begin{figure}[H]
    \centering
    \includegraphics[width=1.0\linewidth]{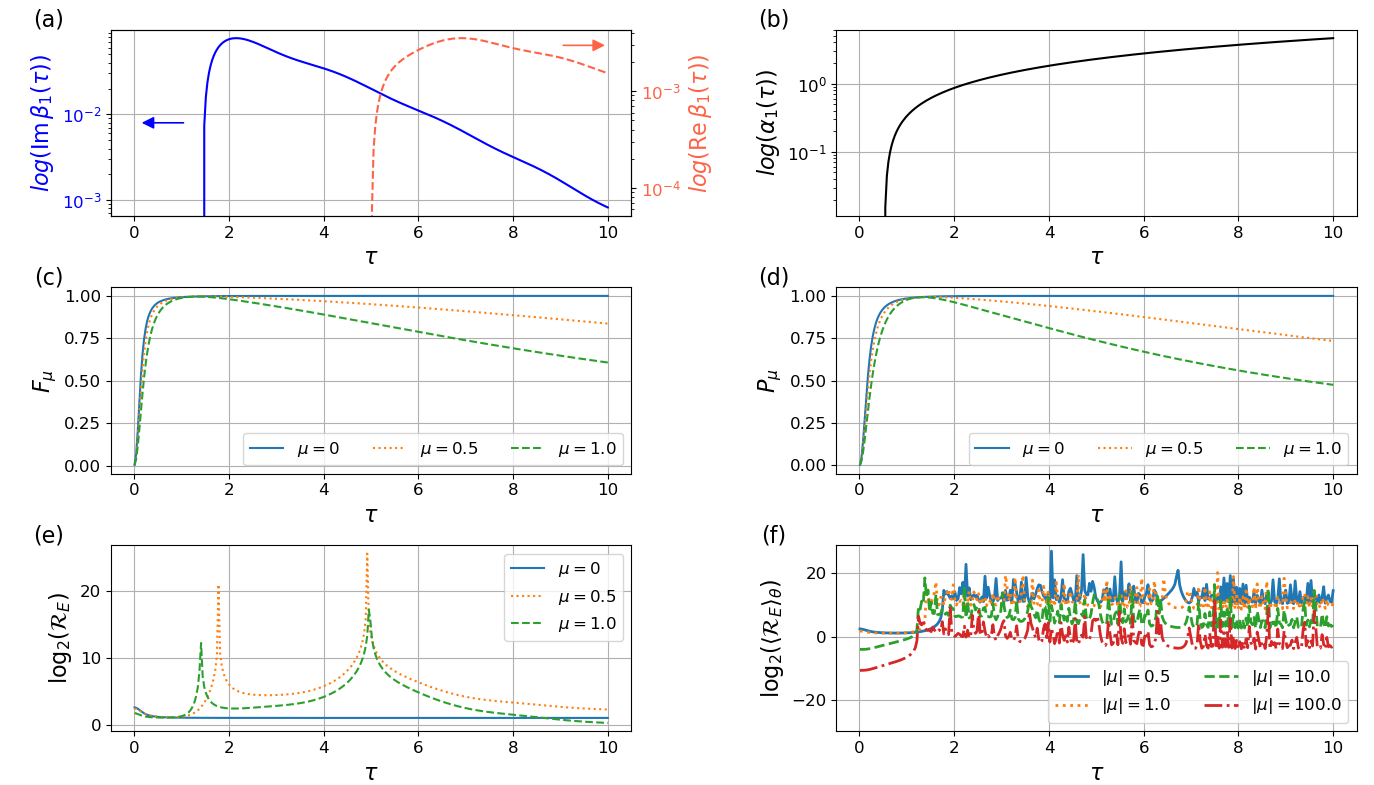}
    \caption{Figures of merit for the finite-time protocol as functions of the protocol duration $\tau$. (\textbf{a}) Logarithm of the real and imaginary parts of the Bogoliubov coefficient $\beta_1$. (\textbf{b}) Logarithm of $\alpha_1$. These quantities characterize parametric excitations produced by the driving. (\textbf{c}) Fidelity with the target STA state. (\textbf{d}) Purity of the reduced oscillator state. Both quantifying precision and irreversibility induced by discarding the clock degree of freedom. (\textbf{e}) Logarithm of the thermodynamic-uncertainty-type ratio $\mathcal{R}_E$ defined in Eq.~\eqref{eq:RE_def} for different coherent-state amplitudes $\mu$. (\textbf{f}) Phase-averaged ratio as a function of $\left| \mu \right|$. These results illustrate the crossover between strong nonadiabatic effects at short durations and suppressed excitations in the slow-driving regime.}
    \label{Fig:finite_time_plots}
\end{figure}

The figures of merit and TUR ratio for the finite time protocol are plotted in Figure~\ref{Fig:finite_time_plots}. For the $\alpha_1$ and $\beta_1$ coefficients, we observe the typical $\sim \frac{1}{\tau}$ behavior for short $\tau$. For long $\tau$ we observe the other typical behavior, $\sim e^{-c_1 \tau}$ for $\beta_1$ (real and imaginary parts) and $\sim e^{c_2 \tau}$ for $\alpha_1$, with $c_1$ and $c_2$ positive constants. For the coherent states, where $\mu \neq 0$, both purity and fidelity reach a maximum for a critical $\tau$ and then decrease monotonically. The case where $\mu = 0$, the vacuum, tends asymptotically to a maximum value. The TUR ratio $\mathcal{R}_E$ is a complicated function that depends not only on $\left|\mu\right|$ and $\tau$, but also on the phase of $\mu$ and has some resonances where particle absorption or emission occurs. Nevertheless, it has a strictly positive minimum that is asymptotically decreasing on its $\left|\mu\right|$ dependence. Given that controlling the exact phase of the initial state can be challenging, we believe that the physically meaningful quantity is the average, so we computed the $\mu$-phase averaged TUR ratio $\mathcal{R}_E$, where the decreasing $\left| \mu \right|$ dependence is clear (see Figure~\ref{Fig:finite_time_plots}).

%Top left: logarithm of the real and imaginary parts of $\beta_1$. Top right: logarithm of $\alpha_1$. Center left: fidelity. Center right: purity. Bottom left: logarithm of the TUR ratio for different values of $\mu$. Bottom right: logarithm of the phase-averaged TUR ratio for different values of $\left|\mu\right|$.

\subsection{Infinite time protocol}
\begin{figure}[H]
    \centering
    \includegraphics[width=0.7\linewidth]{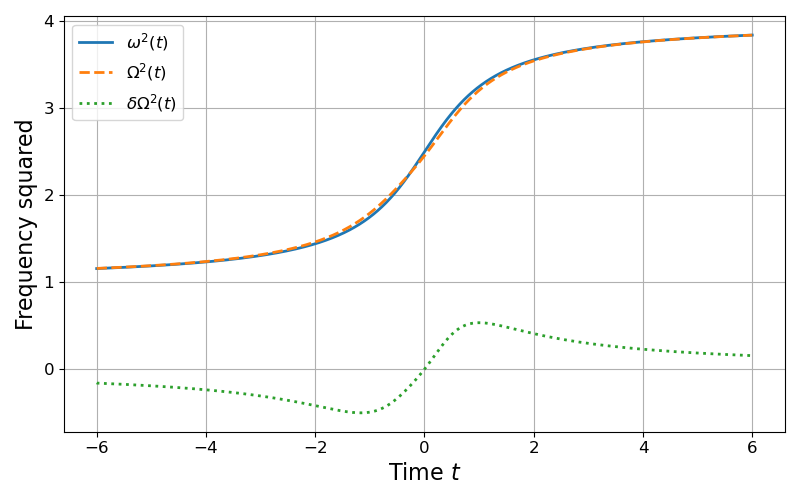}
    \caption{Frequency protocols for the infinite-time driving schedule defined in Eq.~\eqref{eq:inf_prot}. The blue solid curve shows a time window of the bare frequency squared $\omega^2\left(t\right)$, which asymptotically approaches constant values at early and late times. The orange dashed curve shows the shortcut-to-adiabaticity frequency $\bar\Omega^2\left(t\right)$ obtained from Eq.~\eqref{eq:Omegabar}, while the green dotted curve shows the corresponding STA deviation $\delta\Omega^2$ as defined in Eq.~\eqref{eq:deltaOmega_main}. As in Fig.~\ref{Fig:finite_time}, the STA protocol modifies the instantaneous frequency so that the system reproduces the adiabatic mapping in finite time. Frequencies are plotted in units where $\omega_i = 1$ and $\omega_f = 2$, with time measured in units of $\tau$.}
    \label{Fig:infinite_time}
\end{figure}

The standard infinite-time protocol in cosmology is
\be
\label{eq:inf_prot}
\omega^2(t) = \frac{\omega_f^2 + \omega_i^2}{2} + \left( \frac{\omega_f^2 - \omega_i^2}{\pi} \right) \arctan \left( s \right)
\te
with $s = \frac{t}{\tau}$. This protocol begins at $s = -\infty$ and ends at $s = \infty$, satisfying $\lim_{t \to -\infty}\dot{\omega}(t) = \lim_{t \to +\infty}\dot{\omega}(t) = 0$ and $\lim_{t \to -\infty}\ddot{\omega}(t) = \lim_{t \to -\infty}\ddot{\omega}(t) = 0$ and also the Hamiltonian initial and final conditions. The protocol and its STA are plotted in Figure~\ref{Fig:infinite_time}.

\begin{figure}[H]
    \centering
    \includegraphics[width=1.0\linewidth]{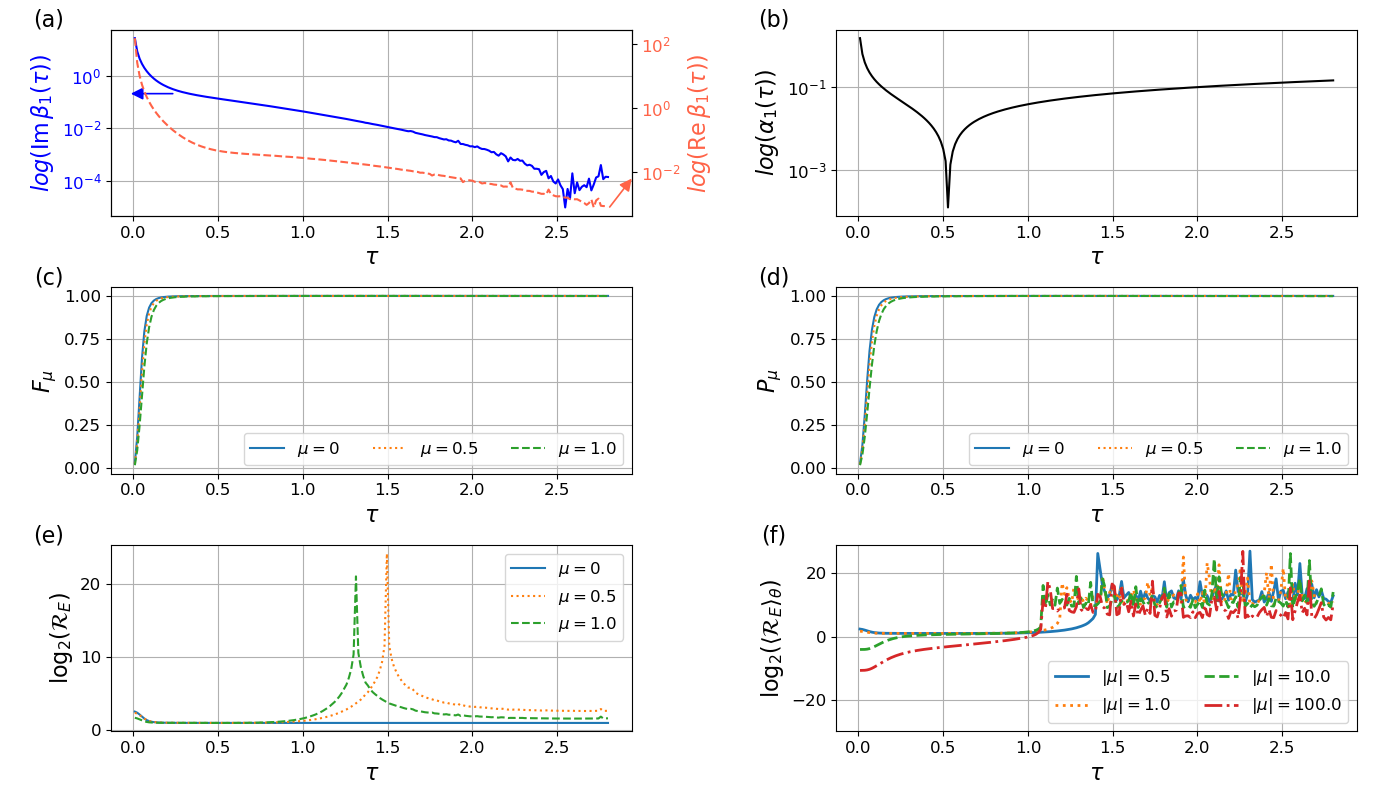}
    \caption{Same quantities as in Fig.~\ref{Fig:finite_time_plots} but for the infinite-time driving protocol defined in Eq.~\eqref{eq:inf_prot}. (\textbf{a}) Logarithm of the real and imaginary part of the (first order expansion) Bogoliubov coefficient $\beta_1$ as a function of the characteristic timescale $\tau$. (\textbf{b}) Logarithm of $\alpha_1$ as a function of the characteristic timescale $\tau$. (\textbf{c}) Fidelity with the target STA state. (\textbf{d}) Purity of the reduced oscillator state. (\textbf{e}) Thermodynamic-uncertainty-type ratio $\mathcal{R}_E$ for different values of the adiabatic parameter $\mu$. (\textbf{f}) Averaged counterpart in logarithmic scale. For short times the dynamics exhibits strong non-adiabatic excitations characterized by $\left|\beta_1\right| \sim \frac{1}{\tau}$, while increasing $\tau$ suppresses excitations and improves fidelity. The TUR ratio remains bounded from below, illustrating the trade-off between energetic fluctuations and irreversibility induced by tracing out the clock.}
    \label{Fig:infinite_time_plots}
\end{figure}

The figures of merit and TUR ratio for the infinite time protocol are plotted in Figure~\ref{Fig:infinite_time_plots}. For the $\alpha_1$ and $\beta_1$ coefficients, we observe the typical $\sim \frac{1}{\tau}$ behavior for short $\tau$. For long $\tau$ we met numerical stability issues due to the regularization needed in the phase of the oscillatory integrals. Nevertheless, analytical approximations indicate that we should observe an exponential decay for $\beta_1$ (real and imaginary parts) in the long-$\tau$ regime. For the coherent states studied, both purity and fidelity tend asymptotically towards a maximum. The TUR ratio $\mathcal{R}_E$ is again a complicated function that depends not only on $\left|\mu\right|$ and $\tau$, but also on the phase of $\mu$ (see Appendix~\ref{app:TUR}) and has some resonances where particle absorption or emission occurs. Nevertheless, it has a strictly positive minimum that is asymptotically decreasing on its $\left|\mu\right|$ dependence. Given that controlling the exact phase of the initial state can be challenging, we believe that the physically meaningful quantity is the average, so we computed the $\mu$-phase averaged TUR ratio $\mathcal{R}_E$, where the decreasing $\left| \mu \right|$ dependence is clear (see Figure~\ref{Fig:infinite_time_plots}).

\section{Conclusions and outlook}
\label{sec:conclusions}

Time-dependent Hamiltonians are an indispensable effective language for quantum control and finite-time thermodynamic strokes. A fully closed microscopic description, however, must be autonomous: the apparent schedule $H(t)$ is generated by extra degrees of freedom that function as a time reference, and the controlled system becomes correlated with that reference. The central message of this work is that, even when the global dynamics is perfectly unitary and the target protocol is engineered to suppress deterministic nonadiabaticity (as in STA), discarding the clock generically produces an effectively open evolution for the system. This induces unavoidable decoherence and a spread of operational outcomes that are intrinsic to time-programmed dynamics.

Concretely, we studied a parametric harmonic oscillator driven by a STA target and implemented autonomously by coupling the oscillator frequency to a minimal pointer clock, building on Ref.~\cite{NotQuiteFreeSTA2018}. Starting from the unitary system--clock dynamics and tracing out the clock with a Feynman--Vernon influence functional, we identified a noise-dominated regime in which the reduced dynamics at the final time is a \emph{random-unitary} mixture, $\bar\rho=\int dv\,P(v)\,\rho_v$, labelled by a single Gaussian clock parameter $v$. This provides a transparent microscopic origin for an ``error model'' of time-dependent control that does not require adding stochasticity by hand.

Within this reduced description, we quantified three operational consequences of clock-induced system--clock correlations: (i) an energetic deviation from the target STA outcome and its associated fluctuations, (ii) loss of overlap with the target trajectory (fidelity/precision), and (iii) loss of purity (irreversibility) of the clock-traced state. For Gaussian inputs, the symplectic formulation makes these quantities computable in closed form (in a controlled small-noise regime), and it also clarifies the role of the Bogoliubov coefficient $\beta$: for each trajectory, $|\beta_v|^2$ is the parametric excitation (squeezing) generated by the $v$-dependent modulation, while $|\beta_1|$ (from $\beta_v = i v\,\beta_1 + O(v^2)$) is the \emph{protocol susceptibility} to clock-induced timing uncertainty. This gives direct physical meaning to the quantity that appears in the numerical plots.

Our results complement and sharpen several related lines of work. A large literature on STA quantifies energetic and thermodynamic costs of fast driving and the tradeoff between speed and resources \cite{DeffnerCampbell2017}. STAs have also been extensively studied in noisy environments \cite{Delvecchio2021,Yin2022,Ritland2018}. Separately, the quantum nature of control degrees of freedom has long been known to limit idealized control, since quantized controllers (e.g.,\ laser modes) generically entangle with the system and induce errors when discarded \cite{vanEnkKimble2002,GeaBanacloche2002,SilberfarbDeutsch2004,Milburn2012}. More recent approaches capture imperfect timekeeping by randomizing the duration of an otherwise time-independent operation \cite{XuerebEtAl2023,Neri2025}, and autonomous clock-driven dynamics has been developed in thermodynamic settings \cite{MalabarbaShortKammerlander2015,LautenbacherEtAl2025}. The present work sits at the intersection of these themes, but differs in emphasis: we focus on \emph{genuinely time-programmed Hamiltonian protocols} (here, an STA schedule) and derive the reduced stochastic description from a first-principles system--clock model. This allows us to express precision loss, purity loss, and energetic deviations in a unified way and to formulate a thermodynamic-uncertainty-type tradeoff between energetic fluctuations and irreversibility generated by discarding the clock.

Several extensions are natural. First, we worked in a noise-dominated approximation and neglected the dissipative kernel of the influence functional at leading order. Restoring it would incorporate systematic backaction/friction-like terms and enable a more complete non-Markovian reduced dynamics; in that setting, it is natural to revisit stochastic representations and their connection to unraveling-type descriptions. Second, while we provided analytic control for Gaussian inputs and small noise, the mixture $\bar\rho$ is generically non-Gaussian; extending the analysis to non-Gaussian resources (e.g.,\ Schr\"odinger-cat superpositions) is a natural next step, both conceptually and for quantum-information applications. Third, our minimal clock can be generalized to more realistic clocks \cite{Viotti2026} (finite bandwidth, multi-mode, or explicitly thermodynamic clocks), and to regimes where the system significantly perturbs the clock, connecting with the growing literature on constrained clock-driven dynamics \cite{LautenbacherEtAl2025}. 

Beyond the present model, the general principle---that autonomous time dependence necessarily produces system--clock correlations and therefore limits idealized time-programmed unitaries---is directly relevant to high-precision coherent control and quantum information processing. Recent work has highlighted thermodynamic constraints on information gain and error correction \cite{DanageozianWildeBuscemi2022}; our results provide a complementary, clock-induced mechanism by which irreversibility and fluctuations arise even without external environments, suggesting a route to principled performance bounds for time-scheduled quantum operations.

\section{Acknowledgements}
This work is funded by the European Union (EQC, 101149233) and supported in part by Universidad de Buenos Aires through Grant No. UBACYT 20020170100129BA, CONICET Grant No. PIP2017/19:11220170100817CO, and ANPCyT Grant No. PICT 2018: 03684.

% ============================================================
%  APPENDICES
%============================================================

\appendix

\section{Influence functional for the system--clock model}
\label{app:influence-functional}

In this Appendix, we derive the influence functional representation used in Sec.~\ref{sec:model}.

\subsection{Path-integral representation of the reduced state}

The total Hamiltonian is
\be
H=\frac{p^2}{2m}+\frac{m}{2}\Omega^2[X]\,x^2+\frac{P^2}{2M}.
\tag{\ref{eq:Htot}}
\te
Assuming an initially factorized state $\rho_{\rm tot}(t_i)=\rho_S(t_i)\otimes\rho_C(t_i)$, the reduced density matrix of the oscillator at $t_f$ admits the (exact) double path-integral representation
\bea
\rho_S(x_f,x_f',t_f)
&=& \int dx_i\,dx_i'\,dX_i\,dX_i'\,dX_f
\int_{\tiny{\begin{array}{c}
x(t_i)=x_i\\ x'(t_i)=x_i'\\ X(t_i)=X_i\\ X'(t_i)=X_i'
\end{array}}}^{\tiny{\begin{array}{c}
x(t_f)=x_f\\ x'(t_f)=x_f'\\ X(t_f)=X_f\\ X'(t_f)=X_f
\end{array}}}
Dx\,Dx'\,DX\,DX'
\nn
&&\times
\exp\!\left\{
\frac{i}{\hbar}\Big(S_{SC}[x,X]-S_{SC}[x',X']\Big)
\right\}
\rho_C(X_i,X_i',t_i)\,\rho_S(x_i,x_i',t_i),
\label{eq:rhoS_fullPI}
\tea
where the system--clock action is
\bea
S_{SC}[x,X]=\int_{t_i}^{t_f} dt\left[
\frac{M}{2}\dot X^2+\frac{m}{2}\dot x^2-\frac{m}{2}\Omega^2[X]\,x^2
\right].
\label{eq:SSC}
\tea

\subsection{Expansion around a mean clock trajectory}

We expand the clock history around a prescribed mean trajectory,
\be
X(t)=\bar X(t)+\xi(t),
\tag{\ref{eq:Xsplit}}
\te
and retain the leading correction in $\xi$. Writing
\be
\Omega^2[X]\simeq \Omega^2[\bar X]+\left.\frac{d\Omega^2}{dX}\right|_{\bar X}\xi
\equiv \bar\Omega^2(t)+V(t)\,\xi(t),
\tag{\ref{eq:linearresponse}}
\te
with $V(t)$ given by Eq.~\eqref{eq:Vdef}, the interaction term becomes
\be
S_I[x,\xi]=-\frac{m}{2}\int_{t_i}^{t_f} dt\; V(t)\,\xi(t)\,x^2(t).
\label{eq:SI}
\te

Tracing out the clock degrees of freedom yields the Feynman--Vernon influence functional,
\be
e^{\frac{i}{\hbar}S_{IF}[x,x']}=
\int D\xi\,D\xi'\;e^{\frac{i}{\hbar}\left(S_C[\xi]-S_C[\xi']\right)}
e^{\frac{i}{\hbar}\left(S_I[x,\xi]-S_I[x',\xi']\right)}\,\rho_C(\xi_i,\xi_i',t_i),
\label{eq:IF_def}
\te
where $S_C[\xi]=\int dt\,\frac{M}{2}\dot\xi^2$ for the free pointer clock.

Because the clock is Gaussian (free) and the coupling is linear in $\xi$, $S_{IF}$ is quadratic in the system coordinates and can be written as
\be
S_{IF}[x,x']=S_D[x,x']+iS_N[x,x'].
\tag{\ref{eq:SIFsplit}}
\te
A convenient explicit form is
\bea
S_D[x,x']
&=&
\frac{m^2}{8}\int_{t_i}^{t_f}dt\;V(t)\,\big(x^2-x'^2\big)(t)
\int_{t_i}^{t_f}dt'\;V(t')\,D(t,t')\,\big(x^2+x'^2\big)(t'),
\label{eq:SD_def}
\\
S_N[x,x']
&=&
\frac{m^2}{16}\int_{t_i}^{t_f}dt\;V(t)\,\big(x^2-x'^2\big)(t)
\int_{t_i}^{t_f}dt'\;V(t')\,N(t,t')\,\big(x^2-x'^2\big)(t'),
\label{eq:SN_def}
\tea
with the kernels
\bea
D(t,t')&=&i\langle[\xi(t),\xi(t')]\rangle\,\theta(t-t'),
\label{eq:Dkernel_def}
\\
N(t,t')&=&\langle\{\xi(t),\xi(t')\}\rangle .
\label{eq:Nkernel_def}
\tea
The averages are taken in the initial clock state $\rho_C(t_i)$.

\section{Clock correlators and kernels}
\label{app:clock-kernels}

In this Appendix, we compute the kernels $D(t,t')$ and $N(t,t')$ for the free pointer clock used in the main text.

\subsection{Free pointer evolution}

For the free clock Hamiltonian $H_C=P^2/2M$, the Heisenberg evolution of the pointer fluctuation is
\be
\xi(t)=\hat\xi+\frac{\hat P}{M}\,t.
\tag{\ref{eq:xiHeis}}
\te
Using $[\hat\xi,\hat P]=i\hbar$, we find
\bea
[\xi(t),\xi(t')]&=&\left[\hat\xi+\frac{\hat P}{M}t,\hat\xi+\frac{\hat P}{M}t'\right]
=\frac{i\hbar}{M}(t-t'),
\\
\{\xi(t),\xi(t')\}
&=&
2\hat\xi^2+\frac{t+t'}{M}\,(\hat\xi\hat P+\hat P\hat\xi)+\frac{2}{M^2}\hat P^2\,t\,t'.
\tea
For clock states that are parity-symmetric in $\xi$ (or more generally for which $\langle \hat\xi\hat P+\hat P\hat\xi\rangle=0$), this reduces to
\bea
\langle[\xi(t),\xi(t')]\rangle &=& \frac{i\hbar}{M}(t-t'),
\\
\langle\{\xi(t),\xi(t')\}\rangle
&=&2\langle\hat\xi^2\rangle+\frac{2}{M^2}\langle\hat P^2\rangle\,t\,t'.
\tea
Substituting into Eqs.~\eqref{eq:Dkernel_def} and \eqref{eq:Nkernel_def} yields the kernels quoted in the main text,
\bea
D(t,t')&=&\frac{\hbar}{M}(t-t')\,\theta(t-t'),
\\
N(t,t')&=&2\langle\hat\xi^2\rangle+\frac{2}{M^2}\langle\hat P^2\rangle\,t\,t'.
%\tag{\ref{eq:Nkernel_main}}
\tea

\section{Noise-dominated limit and Hubbard--Stratonovich reduction}
\label{app:hs-reduction}

Here we provide the steps leading from the influence functional to the single-parameter random-unitary form used in Sec.~\ref{subsec:singleparameter}.

\subsection{Factorization of the noise functional}

Using the $t t'$ contribution in $N(t,t')$,
\be
N(t,t')\simeq \frac{2}{M^2}\langle\hat P^2\rangle\,t\,t' \equiv 2\sigma_v^2\,t\,t',
\qquad
\sigma_v^2=\frac{\langle\hat P^2\rangle}{M^2},
\label{eq:N_ttprime}
\te
the noise functional \eqref{eq:SN_def} becomes
\bea
S_N[x,x']
&\simeq&
\frac{m^2}{16}\int dt\;V(t)\,\Delta(t)
\int dt'\;V(t')\,\big(2\sigma_v^2 t t'\big)\,\Delta(t')
\nn
&=&
\frac{m^2}{8}\sigma_v^2
\left[\int dt\; t\,V(t)\,\Delta(t)\right]^2,
\qquad
\Delta(t)\equiv x^2(t)-x'^2(t),
\label{eq:SN_factorized_app}
\tea
which is Eq.~\eqref{eq:SNfactorized_main} (up to the overall $1/\hbar$ factor in $e^{\frac{i}{\hbar}S_{IF}}$).

\subsection{Hubbard--Stratonovich representation}

We use the Gaussian identity
\bea
e^{-\frac{1}{2}\sigma_v^2 I^2}
=\int dv\;P(v)\,e^{i v I},
\qquad
P(v)=\frac{1}{\sqrt{2\pi\sigma_v^2}}\,e^{-\frac{v^2}{2\sigma_v^2}},
\label{eq:HSidentity}
\tea
which matches Eq.~\eqref{eq:SIF_HS_main} in the main text with an explicit normalization.
With
\be
I[x,x']=\frac{m}{2\hbar}\int_{t_i}^{t_f} dt\; t\,V(t)\,\big(x^2(t)-x'^2(t)\big),
\label{eq:Idef}
\te
the noise contribution satisfies
\be
e^{-\frac{1}{2}\sigma_v^2 I^2}=\int dv\,P(v)\,e^{i v I}.
\te
Neglecting the dissipative part $S_D$ at leading order (noise-dominated approximation), we obtain
\bea
e^{\frac{i}{\hbar}S_{IF}[x,x']}
\simeq
\int dv\,P(v)\,
\exp\!\left[-\frac{i m v}{2\hbar}\int_{t_i}^{t_f} dt\; t\,V(t)\,\big(x^2(t)-x'^2(t)\big)\right],
\label{eq:SIF_HS_app}
\tea
which is Eq.~\eqref{eq:SIF_HS_main}.

\subsection{Random-unitary representation}

For each fixed $v$, the exponent in Eq.~\eqref{eq:SIF_HS_app} amounts to modifying the system action by a $v$-dependent quadratic term. This is equivalent to a unitary oscillator evolution with effective frequency
\be
\Omega^2(t)=\bar\Omega^2(t)+v\,\delta\Omega^2(t),
\qquad
\delta\Omega^2(t)=t\,V(t),
\label{eq:Omega_stochastic_app}
\te
leading to the random-unitary channel representation
\be
\rho_S(t_f)\simeq \int dv\,P(v)\;U_v(t_f,t_i)\,\rho_S(t_i)\,U_v^\dagger(t_f,t_i).
%\tag{\ref{eq:randomunitarymap}}
\te
This decomposition makes clear that the reduced dynamics is a mixture of unitary trajectories (labeled by $v$); mixedness arises only after averaging over $v$, i.e.\ after discarding the clock degrees of freedom.

\section{Perturbative construction of the symplectic map $S_v$}
\label{app:Sv_perturb}

In this Appendix, we show how to expand $S_v(t_f)$ in $v$ in a way that respects the symplectic constraint
$S_v^{\mathsf T}JS_v=J$ order by order.

\subsection{Interaction-picture (Dyson) expansion}

Write
\be
A_v(t)=A_0(t)+vA_1(t),
\qquad
A_1(t)=\begin{pmatrix}0&0\\-m\,\delta\Omega^2(t)&0\end{pmatrix}.
\label{eq:A0A1_app}
\te
Let $S_0(t)$ be the fundamental matrix for $v=0$:
\be
\dot S_0(t)=A_0(t)S_0(t),\qquad S_0(t_i)=\mathbb I_2.
\label{eq:S0_def_app}
\te
Define the interaction-picture map
\be
Y_v(t)\equiv S_0^{-1}(t)\,S_v(t),\qquad Y_v(t_i)=\mathbb I_2.
\label{eq:Yv_def_app}
\te
Then, $Y_v$ satisfies
\be
\dot Y_v(t)=v\,B(t)\,Y_v(t),
\qquad 
B(t)\equiv S_0^{-1}(t)A_1(t)S_0(t).
\label{eq:Yv_ODE_app}
\te
Because $B^{\mathsf T}J+JB=0$, one has $B(t)\in\mathfrak{sp}(2,\mathbb R)$ and hence $Y_v(t)\in Sp(2,\mathbb R)$ for all
$v,t$. The Dyson expansion gives
\bea
Y_v(t_f)
&=&\mathbb I_2
+ v\!\int_{t_i}^{t_f}\!dt_1\,B(t_1)
+ v^2\!\int_{t_i}^{t_f}\!dt_1\!\int_{t_i}^{t_1}\!dt_2\,B(t_1)B(t_2)
+O(v^3).
\label{eq:Yv_Dyson_app}
\tea
Finally,
\be
S_v(t_f)=S_0(t_f)\,Y_v(t_f),
\label{eq:Sv_from_Y_app}
\te
which provides explicit integral expressions for the $v$-derivatives of $S_v$ at $v=0$.

\subsection{Equivalent expansion via fundamental solutions}

Alternatively, expand the fundamental solutions of $\ddot y+\Omega_v^2(t)y=0$ as
$u_v=u_0+v u_1+v^2u_2+\cdots$ and $w_v=w_0+v w_1+v^2w_2+\cdots$. Matching powers of $v$ yields the driven hierarchy
\bea
\ddot u_0+\bar\Omega^2u_0&=&0,\qquad u_0(t_i)=1,\ \dot u_0(t_i)=0,\label{eq:u0_app}\\
\ddot u_1+\bar\Omega^2u_1&=&-\delta\Omega^2\,u_0,\qquad u_1(t_i)=\dot u_1(t_i)=0,\label{eq:u1_app}\\
\ddot u_2+\bar\Omega^2u_2&=&-\delta\Omega^2\,u_1,\qquad u_2(t_i)=\dot u_2(t_i)=0,\label{eq:u2_app}
\tea
(and similarly for $w_k$ with $w_0(t_i)=0,\dot w_0(t_i)=1$ and $w_{1,2}(t_i)=\dot w_{1,2}(t_i)=0$).
Substituting into
\be
S_v(t)=
\begin{pmatrix}
u_v(t) & \frac{1}{m}w_v(t)\\[1mm]
m\dot u_v(t) & \dot w_v(t)
\end{pmatrix}
\label{eq:Sv_uw_app}
\te
yields the perturbative symplectic map.
The Wronskian identity $u_v\dot w_v-\dot u_v w_v=1$ ensures symplecticity (up to the truncation order).

% ============================================================
%  APPENDIX: GAUSSIAN AVERAGES AND v-INTEGRALS (TECHNICAL)
% ============================================================

\section{Gaussian clock averages: energy, purity, and HS fidelity}
\label{app:gaussian_averages}

This Appendix collects the technical steps leading to the closed expressions
\eqref{eq:P_closed_main}--\eqref{eq:TUR_final_main}.

% ------------------------------------------------------------
\subsection{Moment propagation for fixed $v$}
% ------------------------------------------------------------

For each fixed $v$, the unitary $U_v$ induces a symplectic map $S_v$ on $\bm R=(x,p)^{\mathsf T}$,
\be
\bm R(t_f)=S_v\,\bm R(t_i),\qquad S_v\in Sp(2,\mathbb R),
\label{eq:Sv_app_def}
\te
hence a Gaussian input remains Gaussian. Writing the input mean vector and covariance at $t_i$ as
\be
\bm d_i\equiv \langle \bm R(t_i)\rangle,\qquad 
V_i\equiv \frac12\langle \{\Delta\bm R(t_i),\Delta\bm R(t_i)^{\mathsf T}\}\rangle,
\qquad \Delta\bm R\equiv \bm R-\bm d,
\label{eq:diVi_app}
\te
the output moments for realization $v$ are
\be
\bm d_v=S_v\bm d_i,\qquad V_v=S_v V_i S_v^{\mathsf T}.
\label{eq:moments_v_app}
\te

We work perturbatively at fixed $t_f$ and write
\be
S_v=S_0+vS^{(1)}+\frac{v^2}{2}S^{(2)}+O(v^3),
\label{eq:Sv_expand_app}
\te
where $S^{(k)}\equiv \partial_v^k S_v|_{v=0}$. A symplecticity-preserving construction of these derivatives is given in
Appendix~\ref{app:Sv_perturb}.
This induces
\bea
\bm d_v&=&\bm d_0+v\bm d^{(1)}+\frac{v^2}{2}\bm d^{(2)}+O(v^3),\qquad 
\bm d^{(1)}=S^{(1)}\bm d_i,\quad \bm d^{(2)}=S^{(2)}\bm d_i,
\label{eq:d_expand_app}
\\
V_v&=&V_0+vV^{(1)}+\frac{v^2}{2}V^{(2)}+O(v^3),
\label{eq:V_expand_app}
\tea
with
\bea
V^{(1)}&=&S^{(1)}V_i S_0^{\mathsf T}+S_0V_i(S^{(1)})^{\mathsf T},
\label{eq:V1_app}
\\
V^{(2)}&=&S^{(2)}V_i S_0^{\mathsf T}+S_0V_i(S^{(2)})^{\mathsf T}+2S^{(1)}V_i(S^{(1)})^{\mathsf T}.
\label{eq:V2_app}
\tea

% ------------------------------------------------------------
\subsection{Energy: mean deviation and clock-induced variance}
% ------------------------------------------------------------

With $H_f=\frac12\bm R^{\mathsf T}G_f\bm R$ and $G_f=\mathrm{diag}(m\omega_f^2,1/m)$,
a Gaussian state with moments $(\bm d,V)$ satisfies
\be
\langle H_f\rangle=\frac12\,\Tr(G_f V)+\frac12\,\bm d^{\mathsf T}G_f\bm d.
\label{eq:Hf_moments_app}
\te
Expanding $\langle H_f\rangle_v$ using \eqref{eq:d_expand_app}--\eqref{eq:V_expand_app} gives
\bea
\langle H_f\rangle_v
&=&
\langle H_f\rangle_0
+ v\,E^{(1)}+\frac{v^2}{2}\,E^{(2)}+O(v^3),
\label{eq:Ef_expand_app}
\\
E^{(1)}&=&\frac12\Tr(G_f V^{(1)})+\bm d_0^{\mathsf T}G_f\bm d^{(1)},
\label{eq:E1_app}
\\
E^{(2)}&=&\frac12\Tr(G_f V^{(2)})
+(\bm d^{(1)})^{\mathsf T}G_f\bm d^{(1)}
+\bm d_0^{\mathsf T}G_f\bm d^{(2)}.
\label{eq:E2_app}
\tea
Since $P(v)$ is even, $\int dv\,P(v)\,v=0$ and $\int dv\,P(v)\,v^2=\sigma_v^2$, hence
\be
\overline{\Delta E}
=\int dv\,P(v)\big(\langle H_f\rangle_v-\langle H_f\rangle_0\big)
=\frac{\sigma_v^2}{2}\,E^{(2)}+O(\sigma_v^4).
\label{eq:DeltaEbar_app}
\te
The clock-induced variance across trajectories (defined in \eqref{eq:sigmaE_def}, which resembles a constancy \cite{PhysRevLett.120.190602}) reads
\be
\mathrm{Var}_v(H_f)=\sigma_v^2\,(E^{(1)})^2+O(\sigma_v^4),
\label{eq:Varv_app}
\te
because the mean shift is $O(\sigma_v^2)$ while the leading fluctuations come from the linear term.

% ------------------------------------------------------------
\subsection{Gaussian overlap kernel}
% ------------------------------------------------------------

For one-mode Gaussian states $\rho_1,\rho_2$ with moments $(\bm d_1,V_1)$ and $(\bm d_2,V_2)$ in the convention $[x,p]=i\hbar$,
their Hilbert--Schmidt overlap is
\be
\Tr(\rho_1\rho_2)
=
\frac{\hbar}{\sqrt{\det(V_1+V_2)}}
\exp\!\left[
-\frac12(\bm d_1-\bm d_2)^{\mathsf T}(V_1+V_2)^{-1}(\bm d_1-\bm d_2)
\right].
\label{eq:HS_overlap_app}
\te
Define the kernel $K(v,v')\equiv \Tr(\rho_v\rho_{v'})$. Then
\be
\mathcal P=\Tr(\bar\rho_S^2)=\iint dv\,dv'\,P(v)P(v')\,K(v,v'),
\qquad
F_{\mathrm{HS}}=\frac{1}{\mathcal P_0}\int dv\,P(v)\,K(v,0),
\label{eq:P_F_kernel_app}
\te
with $\mathcal P_0\equiv K(0,0)=\Tr(\rho_0^2)=\hbar/\sqrt{\det(2V_0)}$.

% ------------------------------------------------------------
\subsection{Quadratic expansion of $-\log K(v,v')$ and evaluation of the $v$-integrals}
% ------------------------------------------------------------

Introduce $\Sigma(v,v')\equiv V_v+V_{v'}$ and $\Delta\bm d(v,v')\equiv \bm d_v-\bm d_{v'}$.
From \eqref{eq:HS_overlap_app},
\be
\log K(v,v')=\log\hbar-\frac12\log\det\Sigma(v,v')
-\frac12\,\Delta\bm d^{\mathsf T}\Sigma^{-1}\Delta\bm d .
\label{eq:logK_app}
\te
Using \eqref{eq:d_expand_app} and \eqref{eq:V_expand_app}, to second order one needs
\bea
\Sigma(v,v') &=& 2V_0 + (v+v')V^{(1)}+\frac{v^2+v'^2}{2}\,V^{(2)}+O(3),
\label{eq:Sigma_expand_app}
\\
\Delta\bm d(v,v')&=&(v-v')\,\bm d^{(1)}+O(2),
\label{eq:Dd_expand_app}
\tea
and in the displacement term it is sufficient to set $\Sigma^{-1}\to (2V_0)^{-1}$ at this order.

Writing the quadratic expansion in the symmetric form
\be
-\log K(v,v')=
-\log \mathcal P_0
\;+\; b\,(v+v')+\frac12\,a\,(v^2+v'^2)\;-\;c\,v v'\;+\;O(3),
\label{eq:minuslogK_app}
\te
one obtains the trace expressions
\bea
Q &\equiv& V_0^{-1}V^{(1)},\label{eq:Q_app}\\
b&=&\frac14\,\Tr(Q),\label{eq:b_app}\\
a&=&\frac14\,\Tr\!\big(V_0^{-1}V^{(2)}\big)-\frac18\,\Tr(Q^2)
+\frac12\,(\bm d^{(1)})^{\mathsf T}V_0^{-1}\bm d^{(1)},\label{eq:a_app}\\
c&=&\frac18\,\Tr(Q^2)+\frac12\,(\bm d^{(1)})^{\mathsf T}V_0^{-1}\bm d^{(1)}.
\label{eq:c_app}
\tea

Exponentiating \eqref{eq:minuslogK_app} yields the Gaussian approximation to the kernel,
\be
K(v,v')\simeq 
\mathcal P_0\,
\exp\!\left[-b(v+v')-\frac12 a(v^2+v'^2)+c\,v v'\right].
\label{eq:K_gauss_app}
\te
Substituting into \eqref{eq:P_F_kernel_app} reduces $\mathcal P$ to a two-dimensional Gaussian integral and
$F_{\mathrm{HS}}$ to a one-dimensional Gaussian integral, giving the closed forms
\eqref{eq:P_closed_main}--\eqref{eq:FHS_closed_main}.
When $b=0$ (stationary kernel), the exponentials drop out and the results reduce to pure square-root expressions.

% ------------------------------------------------------------
\subsection{Link between $\Delta S_2$ and $\overline{\Delta E}$}
% ------------------------------------------------------------

In the small-noise regime, $\Delta S_2=-\ln(\mathcal P/\mathcal P_0)=a\sigma_v^2+O(\sigma_v^4)$ for $b=0$,
while \eqref{eq:DeltaEbar_app} gives $\overline{\Delta E}=\chi_E\sigma_v^2+O(\sigma_v^4)$ with
\be
\chi_E \equiv \frac12\,E^{(2)}
=\frac14\Tr(G_f V^{(2)})+\frac12(\bm d^{(1)})^{\mathsf T}G_f\bm d^{(1)}
+\frac12\bm d_0^{\mathsf T}G_f\bm d^{(2)}.
\label{eq:chiE_app}
\te
This yields \eqref{eq:DeltaS2_energy_link} in the main text.

% ============================================================
%  APPENDIX (detailed equivalence + vacuum check)
% ============================================================
\section{Bogoliubov--symplectic dictionary and consistency check}
\label{app:equivalence_dictionary}

Here we show explicitly that the Bogoliubov description used in the vacuum/coherent calculations and the symplectic description
used for generic Gaussian inputs are two representations of the same Gaussian unitary for each fixed $v$.

\subsection{From Bogoliubov to a symplectic matrix in the target basis}

At the final time, define the target ladder operator and corresponding quadratures by
\begin{equation}
a_T=e^{iI_f}a_f,\qquad
x=\sqrt{\frac{\hbar}{2m\omega_f}}(a_T+a_T^\dagger),\qquad
p=-i\sqrt{\frac{\hbar m\omega_f}{2}}(a_T-a_T^\dagger),
\label{eq:target_quadratures_app}
\end{equation}
so that $H_f=\hbar\omega_f(a_T^\dagger a_T+\tfrac12)$.
For a given noise realization $v$, the evolved annihilation operator is related to $a_T$ by the Bogoliubov map
\begin{equation}
a(v)=\alpha_v^*\,a_T-\beta_v^*\,a_T^\dagger,
\qquad
a^\dagger(v)=\alpha_v\,a_T^\dagger-\beta_v\,a_T,
\qquad
|\alpha_v|^2-|\beta_v|^2=1.
\label{eq:bogoliubov_target_app}
\end{equation}
Introduce the shorthand $u\equiv \alpha_v^*$ and $w\equiv -\beta_v^*$, so that $a(v)=u\,a_T+w\,a_T^\dagger$.
Then $a_T=u^*a(v)-w\,a^\dagger(v)$ and $a_T^\dagger=-w^*a(v)+u\,a^\dagger(v)$.
Substituting these into Eq.~\eqref{eq:target_quadratures_app} yields
\begin{align}
x
&=\sqrt{\frac{\hbar}{2m\omega_f}}
\Big[(u^*-w^*)\,a(v) + (u-w)\,a^\dagger(v)\Big],
\\
p
&=-i\sqrt{\frac{\hbar m\omega_f}{2}}
\Big[(u^*+w^*)\,a(v) - (u+w)\,a^\dagger(v)\Big].
\end{align}
Now express $a(v),a^\dagger(v)$ back in terms of the \emph{same} quadratures $(x,p)$ in the $v$-basis,
\begin{equation}
a(v)=\sqrt{\frac{m\omega_f}{2\hbar}}\,x+\frac{i}{\sqrt{2\hbar m\omega_f}}\,p,
\qquad
a^\dagger(v)=\sqrt{\frac{m\omega_f}{2\hbar}}\,x-\frac{i}{\sqrt{2\hbar m\omega_f}}\,p,
\label{eq:av_from_xp_app}
\end{equation}
and collect coefficients. One obtains a real linear map $\bm R=(x,p)^{\mathsf T}\mapsto S_v^{(T)}\bm R$ with
\begin{equation}
S_v^{(T)}=
\begin{pmatrix}
\Re(u+w) & \displaystyle \frac{1}{m\omega_f}\Im(u-w)\\[2mm]
-\,m\omega_f\,\Im(u+w) & \Re(u-w)
\end{pmatrix},
\qquad
u=\alpha_v^*,\;\; w=-\beta_v^*.
\label{eq:Sv_dictionary_app}
\end{equation}
Using $|\alpha_v|^2-|\beta_v|^2=1$, it is straightforward to verify that $S_v^{(T)\mathsf T}J S_v^{(T)}=J$ (hence $S_v^{(T)}\in Sp(2,\mathbb R)$).
This establishes the equivalence: specifying $(\alpha_v,\beta_v)$ fixes the symplectic map $S_v^{(T)}$, and conversely.

\subsection{Vacuum state in symplectic formalism}

For the target vacuum $\rho_0=|0\rangle_T\,{}_T\langle 0|$, the target covariance in the $(x,p)$ basis is
\begin{equation}
V_0=\frac{\hbar}{2}
\begin{pmatrix}
\frac{1}{m\omega_f} & 0\\
0 & m\omega_f
\end{pmatrix},
\label{eq:V0_app}
\end{equation}
and for each $v$ the output is a squeezed vacuum with covariance $V_v=S_v^{(T)}V_0 S_v^{(T)\mathsf T}$.
The mean remains zero. Therefore the energy deviation is
\begin{equation}
\langle H_f\rangle_v-\langle H_f\rangle_0
=\frac12\Tr(G_f(V_v-V_0))
=\hbar\omega_f\,|\beta_v|^2,
\label{eq:vac_energy_check_app}
\end{equation}
in agreement with the Bogoliubov result in the main text (obtained by normal ordering $H_f$ in the $a(v)$ basis).

Moreover, the Hilbert--Schmidt overlap between two one-mode Gaussian states depends only on their first and second moments.
For the vacuum case, $\bm d_v=0$ and
\begin{equation}
\Tr(\rho_v\rho_{v'})
=
\frac{\hbar}{\sqrt{\det(V_v+V_{v'})}},
\label{eq:HS_vac_app}
\end{equation}
which, using the perturbative expansion $\beta_v\simeq i v\,\beta_1$ in the noise-dominated regime, yields
\begin{equation}
|\langle 0,v|0,0\rangle_T|^2\simeq \exp\!\left[-\tfrac12\,|\beta_1|^2 v^2\right],
\qquad
|\langle 0,v|0,v'\rangle|^2\simeq \exp\!\left[-\tfrac12\,|\beta_1|^2 (v-v')^2\right],
\label{eq:vac_overlap_check_app}
\end{equation}
which are exactly the Gaussian kernels used in the explicit Bogoliubov overlap calculation.
Performing the $v$ (and $v,v'$) Gaussian averages then reproduces
\begin{equation}
F_0=\frac{1}{\sqrt{1+|\beta_1|^2\sigma_v^2}},
\qquad
\mathcal P_0=\frac{1}{\sqrt{1+2|\beta_1|^2\sigma_v^2}},
\label{eq:vac_F_P_check_app}
\end{equation}
as quoted in Appendix~\ref{app:gaussian_averages}.
The coherent-state case follows similarly: the covariance evolves as in the vacuum, while the mean is transported linearly by $S_v^{(T)}$,
so all observables split into a covariance contribution (squeezing) plus a mean contribution (displacement).

\section{TUR}
\label{app:TUR}
Because of computational difficulties, we formulate our TUR-type relation in terms of the linear-entropy production and compute it for initially coherent states in terms of the Bogoliubov coefficients:
\bea
\mathcal S_L\,\frac{\overline{\sigma_E^2}}{\overline{\Delta E}^{\,2}} &=& \left( \frac{4 \left[ 1 + 2 \left|\mu\right|^2 \left( 1 + 2r\cos{\varphi} + r^2 \right) \right]}{1 + 2 \sigma_v^2 \left|\beta_1\right|^2 \left[ 1 + 2 \left|\mu\right|^2 \left( 1 + 2r\cos{\varphi} + r^2 \right) \right] + \sqrt{1 + 2 \sigma_v^2 \left|\beta_1\right|^2 \left[ 1 + 2 \left|\mu\right|^2 \left( 1 + 2r\cos{\varphi} + r^2 \right) \right]}} \right) \times \nn
&\times& \frac{\left\{ 1 + 3 \left| \beta_1 \right|^2 \sigma_v^2 + 2 \left|\mu\right|^2 \left[ 1 + 6 \left| \beta_1 \right|^2 \sigma_v^2 \left( 1 + r \cos{\varphi} \right) \right] \right\}}{\left[ 1 + 2 \left| \mu \right|^2 \left( 1 + r \cos{\varphi} \right) \right]^2} = \mathcal{R}_E
\tea
where $r := \frac{\alpha_1}{\left|\beta_1\right|}$ and $\cos{\varphi} := \operatorname{Re}\left( \frac{\beta_1}{\left|\beta_1\right|} \Delta\right)$. In order to make it human-readable, we write
\bea
A &\equiv& 1 + 2 r \cos{\varphi} + r^2 = \left(1 + r e^{i\varphi} \right) \left( 1 + r e^{-i\varphi} \right) \nn
B &\equiv& 1 + r \cos{\varphi} \nn
X &\equiv& 2 \sigma_v^2 \left| \beta_1 \right|^2 \nn
Y &\equiv& 1 + 2 \left| \mu \right|^2 A
\tea
and find
\be
\mathcal{R}_E = \frac{4 Y}{1 + X Y + \sqrt{1 + X Y}} \times \frac{1 + \frac{3}{2} X + 2 \left| \mu \right|^2 \left[ 1 + 3 X B \right]}{\left[ 1 + 2\left| \mu \right|^2 B \right]^2}.
\te
The first factor behaves as $\approx 2 Y$ for $X \ll 1$ and $\approx \frac{4Y}{X Y} = \frac{4}{X}$ for $X \gg 1$. The second factor is asymptotically decreasing for large $\left| \mu \right|^2$, but for any fixed value, a strictly positive bound can be found.

%% The reference list below matches the published version (Entropy 2026, 28, 519).
%% If you prefer BibTeX, delete the thebibliography block and uncomment the two lines below,
%% using the accompanying biblio.bib file.
%\bibliographystyle{unsrt}
%\bibliography{biblio}

\end{document}